\documentclass{aa}
\usepackage{amssymb,amsfonts,amsmath}
\usepackage[varg]{txfonts}
\usepackage{natbib}
\usepackage{graphicx}

\usepackage{color}
\definecolor{mygray}{gray}{0.5}
\definecolor{magenta}{rgb}{0.858, 0.188, 0.478}

\usepackage{xspace}
\newcommand{\fg}[1]{Fig.~\ref{fig:#1}}
\newcommand{\Fg}[1]{Figure~\ref{fig:#1}}
\newcommand{\eq}[1]{Eq.~(\ref{eq:#1})}
\newcommand{\se}[1]{Sect.~\ref{sec:#1}}
\newcommand{\Se}[1]{Section~\ref{sec:#1}}

\newcommand{\Tb}[1]{Table~\ref{tab:#1}}

\newcommand{\paplet}{Paper\xspace}
\newcommand{\ie}{\textit{i.e.,}\xspace}
\newcommand{\eg}{\textit{e.g.,}\xspace}
\newcommand{\trappisti}{\mbox{TRAPPIST-1}\xspace}

\newcommand{\Fpg}{\mathcal{F}_\mathrm{p/g}}
\newcommand{\Fsg}{\mathcal{F}_\mathrm{s/g}}
\newcommand{\epa}{\epsilon_\mathrm{PA}}
\newcommand{\hiio}{\mathrm{H}_2\mathrm{O}}

\begin{document}

\title{Formation of \trappisti and other compact systems}

\author{Chris W. Ormel, Beibei Liu, and Djoeke Schoonenberg}

\institute{Anton Pannekoek Institute (API), University of Amsterdam, Science Park 904,1090GE Amsterdam, The Netherlands\label{inst1} \\
\email{[c.w.ormel,b.liu,d.schoonenberg]@uva.nl}
   }

\date{\today}

\abstract{\trappisti is a nearby 0.08\,$M_\odot$ M-star, which was recently found to harbor a planetary system of at least seven Earth-size planets, all within 0.1\,au. The configuration confounds theorists as the planets are not easily explained by either \textit{in situ} or migration models. In this \paplet we present a scenario for the formation and orbital architecture of the \trappisti system. In our model, planet formation starts at the H$_2$O iceline, where pebble-size particles -- whose origin is the outer disk -- accumulate to trigger streaming instabilities. After their formation, planetary embryos quickly mature by pebble accretion. Planet growth stalls at Earth masses, where the planet's gravitational feedback on the disk keeps pebbles at bay. Planets are transported by Type I migration to the inner disk, where they stall at the magnetospheric cavity and end up in mean motion resonances. During disk dispersal, the cavity radius expands and the inner-most planets escape resonance. We argue that the model outlined here can also be applied to other compact systems and that the many close-in super-Earth systems are a scaled-up version of \trappisti. We also hypothesize that few close-in compact systems harbor giant planets at large distances, since they would have stopped the pebble flux from the outer disk.
} 

\maketitle

\section{Introduction}
\trappisti\ -- a late-type 0.08$\,M_\odot$ M-star situated at a distance of 12\,pc -- is known to harbor an ultra-compact planetary system of at least six planets \citep{GillonEtal2016,GillonEtal2017}. The seventh planet, \trappisti{h}, was recently confirmed by a transit timing analysis and \textit{K2} data \citep{LugerEtal2017}. All planets are within 0.1\,au of their host star. Remarkably, all planets have masses similar to Earth and the inferred densities, although highly uncertain, are consistent with rocky compositions \citep{GillonEtal2017}. The orbital period ratios of the \trappisti planets indicate that planets d/e, e/f and g/h are very close to 3:2 mean motion resonance (MMR) while f/g shows a 4:3 commensurability. Planets b/c and c/d are somewhat further located from a first order MMR.

The ultra-compact configuration of the \trappisti system raises the question how was it formed. \textit{In situ} formation, where rocky planets emerge from a giant impact phase \citep[\eg][]{HansenMurray2012}, would have required an unusually dense disk and would also not easily explain the resonant configuration.  Planet migration seems to be a more plausible model \citep{LeePeale2002}. However, formation beyond the iceline cannot explain the predominantly rocky composition. In addition, traditional formation scenarios fail to explain why all planets end up at masses approximately equal to Earth's. 

In this \paplet we hypothesize a different scenario: it was the planetary building blocks in the form of mm/cm-size particles (pebbles) that migrated to the inner disk. 
Circumstellar disks contain large amounts of pebble-size particles \citep[\eg][]{TestiEtal2014,PerezEtal2015}; and the thermal emission from these particles has also been observed around low-mass stars or even brown dwarfs \citep{RicciEtal2012}.  
We argue that these pebbles were transformed into planetary embryos at the H$_2$O iceline, situated at $\approx$0.1\,au, where they migrated inwards by type~I migration. Interior to the H$_2$O iceline the planets grew by accretion of rocky pebbles. Inward planet migration ceased at the disk's inner edge, where they migrated into first-order MMR. Later, during the disk dispersion phase, the 3:2 MMRs of the inner two planet pairs were broken, resulting in the architecture that we witness today.

The goal of this \paplet is to present a synopsis of the early history of the \trappisti system from simple analytical reasoning, which may inspire future numerical, more precise treatments. The adopted disk and stellar parameters (\Tb{params}) have been optimized towards \trappisti but are not out of the ordinary. In \se{disk} we discuss the circumstellar disk of \trappisti. \Se{synopsis} presents the chronology. In \se{summary} we speculate about the implications of our model to other stars.

\section{The \trappisti disk}
\label{sec:disk}
\begin{table*}
    \centering
    \caption{Default disk and stellar parameters of \trappisti during the planet formation phase \label{tab:params}.}
    \begin{tabular}{lp{8cm}ll}
    \hline
    \hline
    symbol  &   description &   value   &   comments \\
    \hline
    $\alpha$                & viscosity parameter inner disk           & $10^{-3}$         & (a) \\
    $\delta_\mathrm{ice}$   & fractional width of iceline              & $0.05$ \\
    $\gamma_I$              & prefactor in Type-I migration rate            & $4$               & (b) \\
    $\xi$                   & \# $e$-foldings to reach pebble size          & 10 \\
    $\tau_p$                & pebble dimensionless stopping time 
                                        exterior to iceline            & 0.05 & (c) \\
    $\zeta$                 & dust fraction in pebbles                 & 0.5 \\
    $h$                     & inner disk aspect ratio                       & $0.03$ \\
    $r_\mathrm{out}$        & outer disk radius                        & $200\,\mathrm{au}$ \\
    $B_\star$               & stellar magnetic field strength at surface    & $180\:\mathrm{G}$ \\
    $M_\star$               & stellar mass                                  & $0.08\:M_\odot$ \\
    $M_\mathrm{disk}$       & disk mass (gas)                          & $0.04\:M_\star$\\
    $\dot{M}_g$             & accretion rate                                & $10^{-10}\:M_\odot\:\mathrm{yr}^{-1}$ \\
    $R_\star$               & stellar radius                                & $0.5\:R_\odot$ \\
    $Z_0$                   & disk metallicity (global solids-to-gas mass ratio)                  & $0.02$ \\
    \hline
    \hline
    \end{tabular}
    \tablefoot{\tablefoottext{a}{Our default model. We also discuss more laminar and more turbulent disks ($10^{-4}\le\alpha\le10^{-2}$)}; \tablefoottext{b}{\citet{KleyNelson2012}}; \tablefoottext{c}{The stopping time of pebbles interior to the iceline is denoted $\tau_s$ with $\tau_s \ll \tau_p$}.}
\end{table*}
\subsection{Disk structure}
We assume that the \trappisti circumstellar disk can be divided in two regions:
\begin{itemize}
    \item The inner disk, $r\ll 1\,\mathrm{au}$. This is the region where the iceline is located, and where the planetary system forms. We further assume that the inner disk is viscously relaxed, characterized by an $\alpha$-viscosity, $\nu=\alpha h^2 r^2 \Omega$, where $h$, the aspect ratio, is assumed constant and $\Omega_K(r)$ the local orbital frequency. The gas surface density $\Sigma_g$ follows from the accretion rate $\dot{M}_g$: $\Sigma_g = \dot{M}_g/3\pi\nu$ \citep{Lynden-BellPringle1974}, where we adopt a value of $\dot{M}_g=10^{-10}\,M_\odot\,\mathrm{yr}^{-1}$ typical for M-stars \citep[\eg][]{ManaraEtal2015}. The constant aspect ratio of the inner disk is motivated by viscous heating and lamppost heating ($R_\star/r\sim0.1$; \citealt{RafikovDe-Colle2006}), as well as from SED-fitting \citep{MuldersDominik2012}. The temperature structure corresponding to $h=0.03$ reads:
\begin{equation}
    T(r) = 180\:\mathrm{K}\: \frac{M_\star}{0.08\:M_\odot} \left( \frac{h}{0.03} \right)^2 \left( \frac{r}{0.1\:\mathrm{au}} \right)^{-1}
\end{equation}
\item The outer disk, $r\gg 1\,\mathrm{au}$ (\fg{outer-disk}). This is the region that dominates the disk mass (in solids as well as gas). Since viscous timescales ($\sim$$r^2/\nu$) are longer than the duration of the planet formation process (see \eq{t-end}) a steady accretion disk is inappropriate. Instead, we simply adopt a power-law profile for the surface density:
    \begin{equation}
        \Sigma_\mathrm{g,out} = \frac{(2-p)M_\mathrm{disk}}{2\pi r_\mathrm{out}^2} \left( \frac{r}{r_\mathrm{out}} \right)^{-p};\qquad (p<2)
        \label{eq:sigma-out}
    \end{equation}
    where $M_\mathrm{disk}$ is the total disk mass and $r_\mathrm{out}$ the disk's outer radius. We choose $p=1$, $M_\mathrm{disk}=0.04\,M_\star$ and a metallicity of $Z_0=0.02$, which amounts to a total mass of $\approx$22 Earth masses in solids of which $\approx$$11\,M_\oplus$ is in rocky material (assuming a dust fraction in pebbles of $\zeta=0.5$; \citealt{Lodders2003}).
\end{itemize}

The inner disk is truncated at the magnetospheric cavity radius \citep[\eg][]{FrankEtal1992}:
\begin{equation}
    \label{eq:rc}
    r_c = \left( \frac{B_\star^4 R_\star^{12}}{4GM_\star \dot{M}_g^2} \right)^{1/7}
    \approx 0.0102\, \mathrm{au}
\end{equation}
where $B_\star$ is the strength of the magnetic field measured at the surface of the star ($R_\star$).  In this and other equations the numerical value follows from inserting the default parameters listed in \Tb{params}. Here, the surface magnetic field strength of 180\,G -- perhaps lower than the typical $\sim$kG of T-Tauri stars -- is consistent with observations of brown dwarfs \citep{ReinersEtal2009}.

\begin{figure}
    \centering
    \includegraphics[width=88mm]{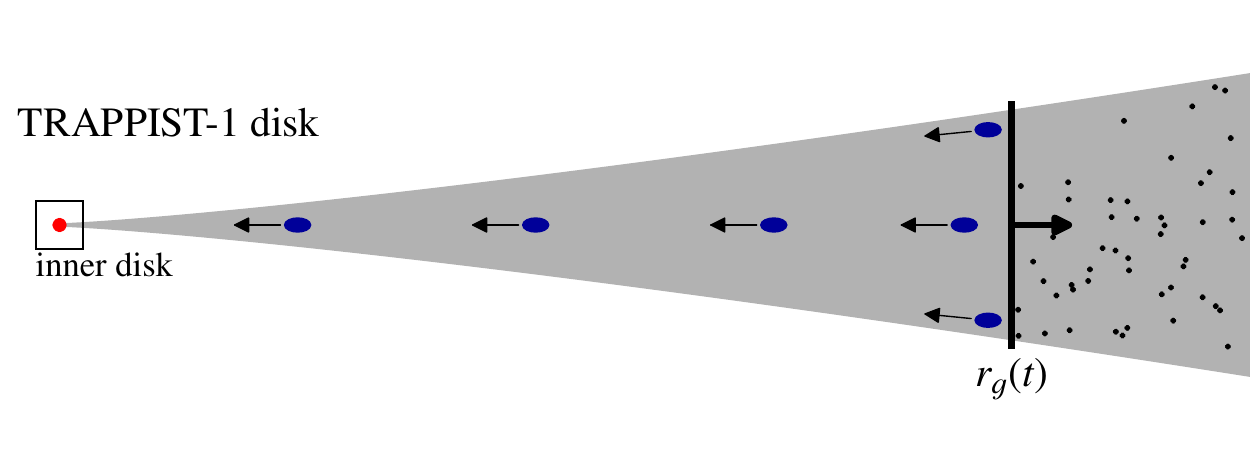}
    \caption{\trappisti outer disk. Dust grains coagulate and drift inwards, resulting in a pebble front $r_g$ that moves outwards with time. After $t=10^5\,\mathrm{yr}$ the pebble front reaches 50\,au. The box surrounding the star is then 1\,au in size.}
    \label{fig:outer-disk}
\end{figure}
In general, disks feature a negative radial pressure gradient, which causes gas to rotate slower than Keplerian by an amount $\eta v_K$ \citep{NakagawaEtal1986}. For our choices of $\Sigma_g(r)$ and $T(r)$ for the inner disk we obtain that $\eta = \frac{5}{4}h^2$ is constant. The sub-Keplerian motion induces the orbital decay of pebble-size particles \citep{Weidenschilling1977}: $v_r = - 2\eta v_K \tau_p/(1+\tau_p^2) \approx -2\eta v_K \tau_p$
where $\tau_p = t_\mathrm{stop} \Omega_K$, the dimensionless stopping time, is assumed less than unity. We assume that just exterior to the iceline $\tau_p=0.05$, a value typical for drifting pebbles \citep{BirnstielEtal2012,LambrechtsJohansen2014,KrijtEtal2016}. For our standard disk parameters $\tau_p=0.05$ at 0.1\,au corresponds to a physical size of 4\,cm. Interior to the iceline the size (and $\tau_p$) will be smaller, because icy pebbles disintegrate.

\begin{figure*}
    \centering
    \includegraphics[width=80mm]{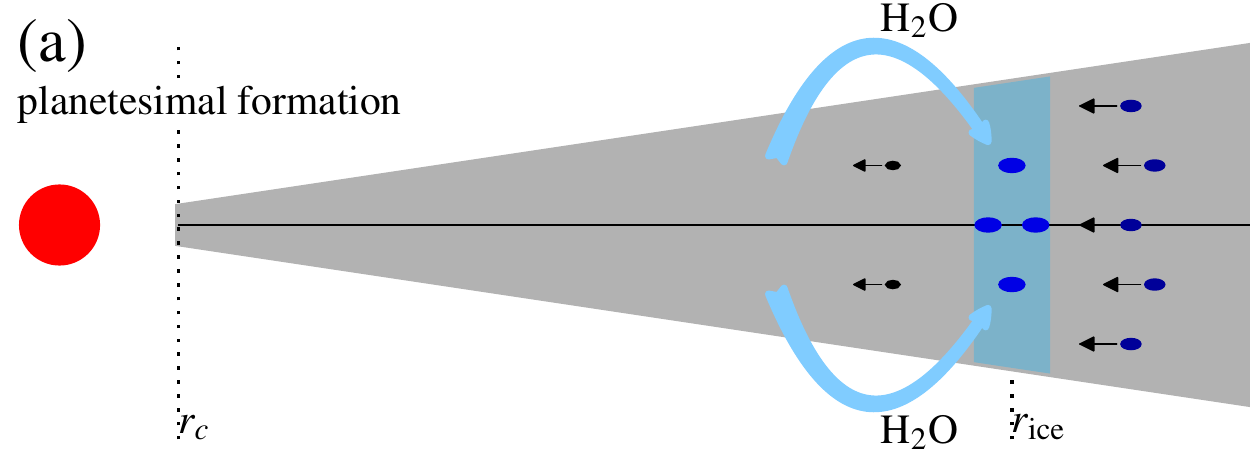}\qquad\qquad\includegraphics[width=80mm]{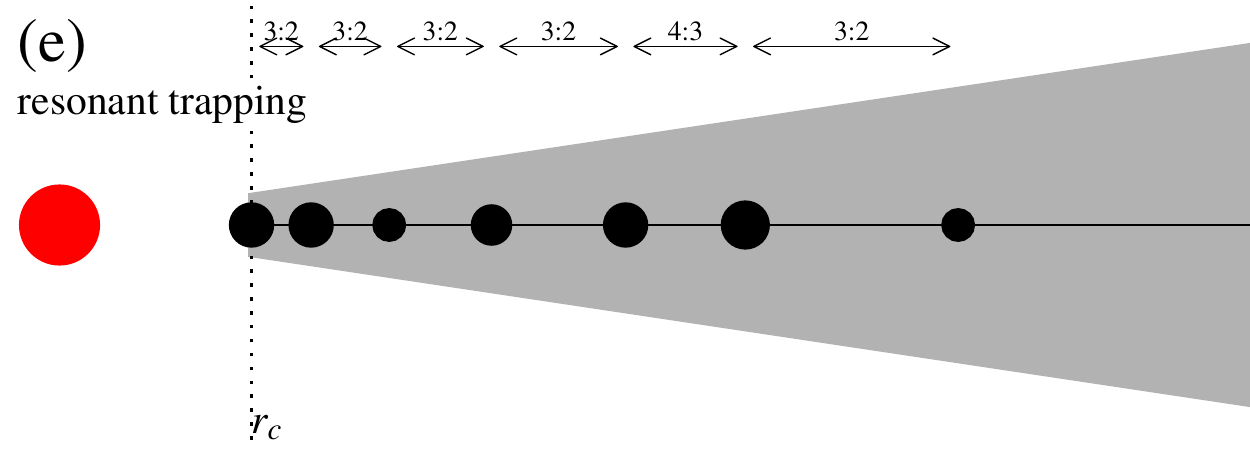}
    \includegraphics[width=80mm]{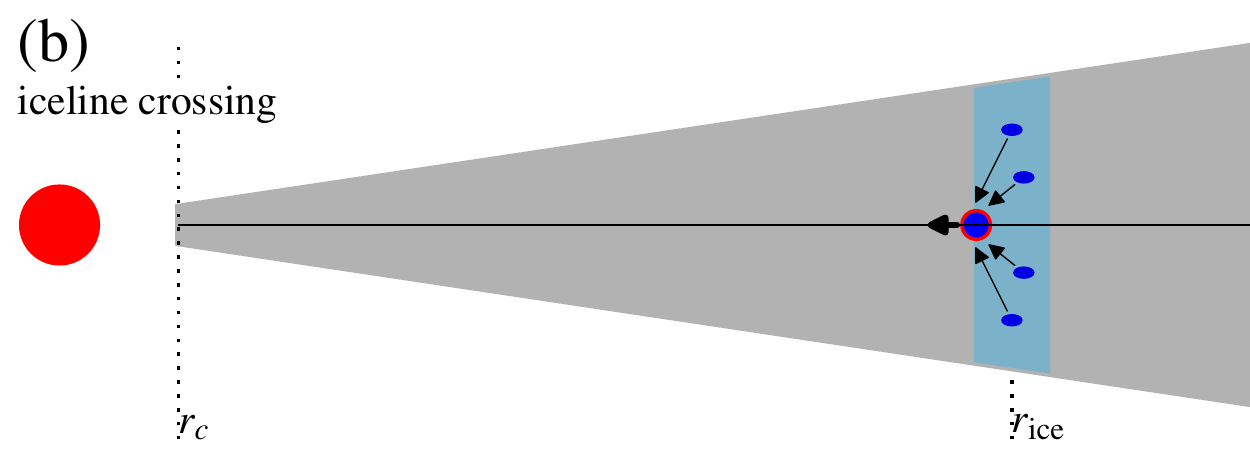}\qquad\qquad\includegraphics[width=80mm]{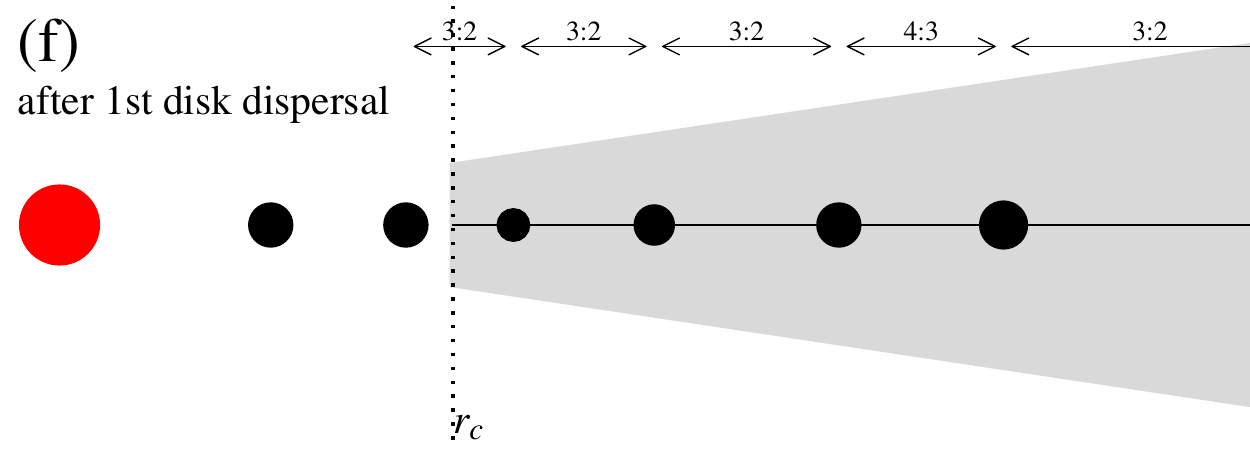}
    \includegraphics[width=80mm]{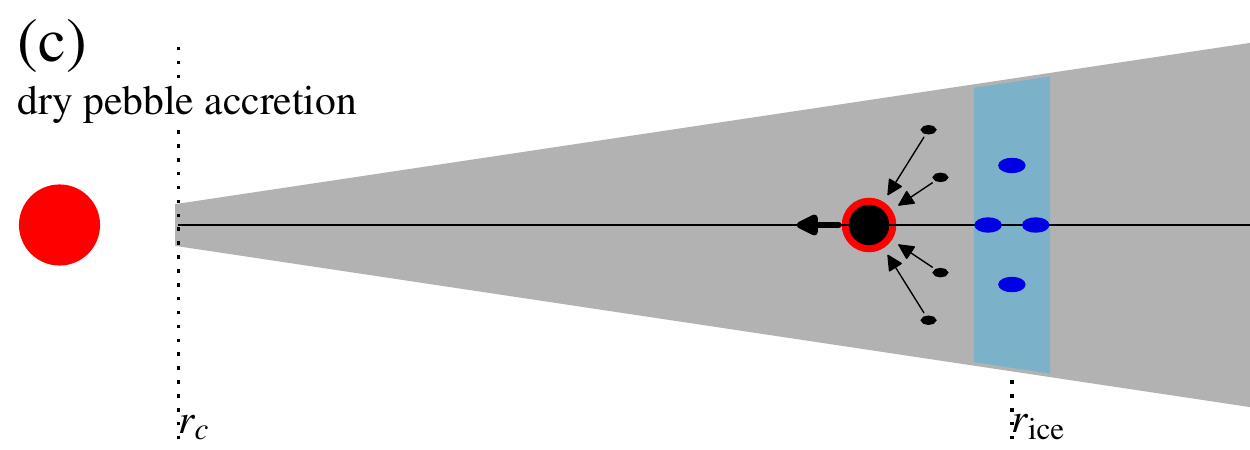}\qquad\qquad\includegraphics[width=80mm]{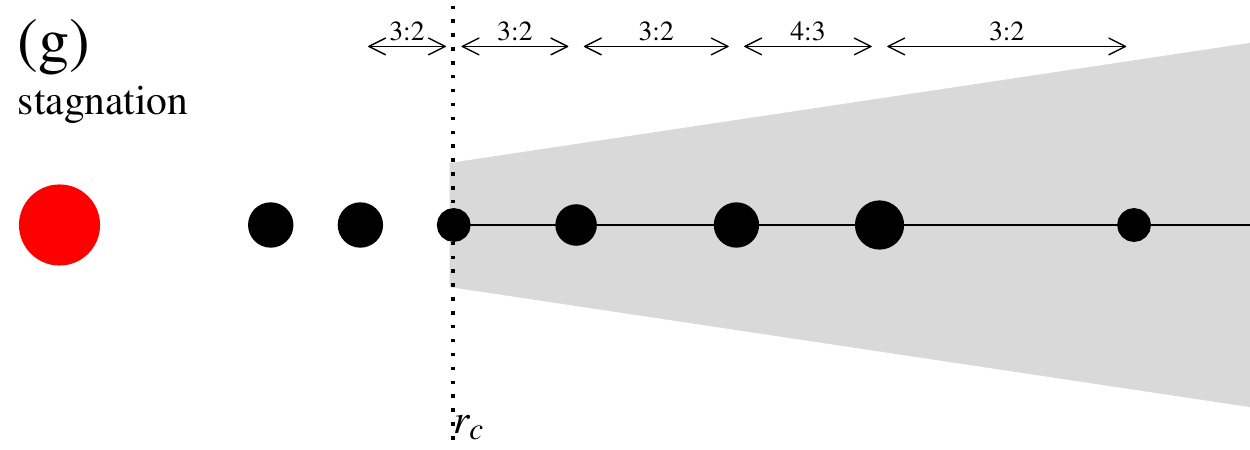}
    \includegraphics[width=80mm]{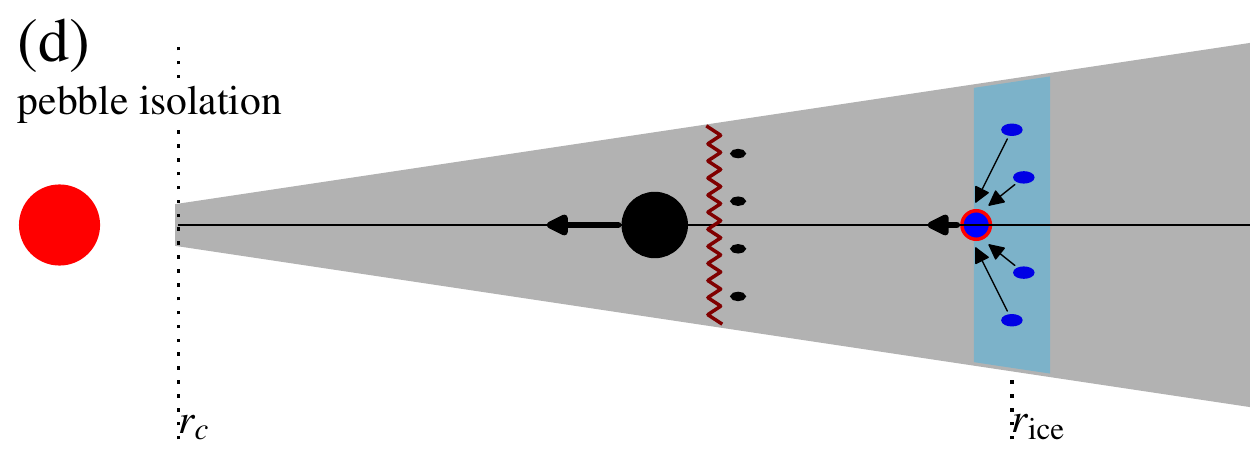}\qquad\qquad\includegraphics[width=80mm]{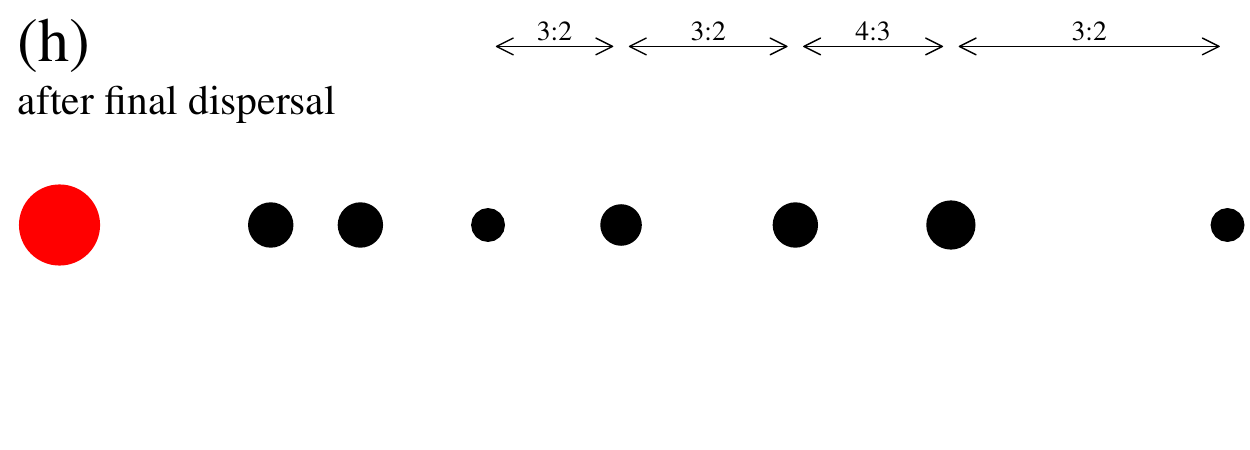}
    \caption{Stages in the formation of the \trappisti system. Panels shows the inner disk ($r\lesssim0.1\,\mathrm{au}$): left panels describe the formation phases; right panels the disk dispersion and dynamical rearrangement. Stages are described in \se{synopsis}. Sizes and aspect ratios are not to scale, but relative distances are.}
    \label{fig:sketch}
\end{figure*}
\subsection{The pebble mass flux}
Most of the solids are located in the outer disk, initially in micron-size grains. Using an $n$-$\sigma$-$\Delta v$ calculation, it follows that the collision timescales among grains is $t_\mathrm{coll} \sim (Z_0\Omega_K)^{-1}$, where $Z_0$ is the initial dust-to-gas ratio. Growth to pebble-size particles -- large enough to start drifting -- involves many binary collisions. Let $t_\mathrm{peb} = \xi(Z_0\Omega_K)^{-1}$, with $\xi\approx10$ \citep{KrijtEtal2016}, be the time needed before drifting pebbles appear. 
Equating $t=t_\mathrm{peb}$ and solving for $r$ we find the radius of the pebble front $r_g$ where the particles reach their drifting size:
\begin{equation}
    r_g = \left( \frac{G M_\star Z_0^2 t^2}{\xi^2} \right)^{1/3} = 50\,\mathrm{au} \left( \frac{t}{10^5\,\mathrm{yr}} \right)^{2/3}. 
    \label{eq:rg}
\end{equation}
The advancing $r_g(t)$ generates pebbles at a rate $\dot{M}_P = 2\pi r_g \dot{r}_g Z_0 \Sigma_\mathrm{g} (r_g)$ and results in a pebble-to-gas mass flux ratio of:
\begin{equation}
    \label{eq:Fpg}
    \Fpg
    \equiv \frac{\dot{M}_p}{\dot{M}_g}
    = \frac{2 M_\mathrm{disk} Z_0^{5/3}}{3\dot{M}_g r_\mathrm{out} \xi^{2/3}} \left( \frac{GM_\star}{t} \right)^{1/3} 
    \approx 1.1 \left( \frac{t}{10^5\,\mathrm{yr}} \right)^{-1/3}
\end{equation}
\citep[cf.][]{LambrechtsJohansen2014}. The pebble flux will disappear after the time $t_\mathrm{end}$, when the pebble front hits the outer edge of the disk, $r_g=r_\mathrm{out}$:
\begin{equation}
    t_\mathrm{end} 
    = \frac{\xi}{Z_0}\sqrt{\frac{r_\mathrm{out}^3}{GM_\star}}
    \approx 8\times10^5\,\mathrm{yr}
    \label{eq:t-end}
\end{equation}

\section{Synopsis}
\label{sec:synopsis}
In our scenario, the formation of the \trappisti planets proceeds in two stages. In the first stage, planetary embryos assemble sequentially at the iceline, migrate inwards, and end up in resonance near the disk edge  (panels a--d). The second stage concerns the dynamical re-arrangement, triggered by the disk dispersal, which moves the inner planets out of MMR (panels e--h).

In the first stage, our assumption is that planets form \textit{sequentially}, not simultaneously. In our model we assume that the H$_2$O iceline is the location where the midplane solids-to-gas ratio exceeds unity, triggering streaming instabilities and spawning the formation of planetesimals. These planetesimals merge into a planetary embryo, whose growth is aided by icy pebble accretion. Once its mass becomes sufficiently large, it migrates interior to the H$_2$O iceline by type~I migration, where it continues to accrete (now dry) pebbles until it reaches the pebble isolation mass.  After some time, a second embryo forms at the snowline, which follows a similar evolutionary path as its predecessor. Even though the inner planet's growth could be reduced by its younger siblings' apetite for pebbles, it always remains ahead in terms of mass. Planet migration stalls at the inner disk edge, where the planets are trapped in resonance.

\subsection{Formation of planetesimals (a)}
The first step is the concentration of pebble-size particles and their subsequent gravitational collapse into planetesimals. 
A prominent mechanism is the streaming instability, where particles clump into filaments because of the backreaction of the solids on the gas \citep{YoudinGoodman2005}. 
These filaments fragment and spawn planetesimals \citep{JohansenEtal2007}. Recent work has demonstrated that streaming instabilities can be triggered for a broad range of stopping times \citep{YangEtal2016}; however a prerequisite is that the solids-to-gas ratio must be substantial \citep{JohansenEtal2009,CarreraEtal2015,YangEtal2016}. Since streaming instabilities arise by virtue of the backreaction of the solids on the gas, we seek volume solids-to-gas ratios $\rho_p/\rho_g\sim$1.

However, the inside-out growth and drift of solids do not guarantee large solids-to-gas ratios \citep{BirnstielEtal2010i,KrijtEtal2016,SatoEtal2016}. The midplane pebble-to-gas density ratio is only 
\begin{equation}
    \left(\frac{\rho_p}{\rho_g}\right)_\mathrm{midplane}
    = \frac{\Sigma_p/H_p}{\Sigma_g/H_g}
    = \frac{3 \Fpg}{5} \sqrt{\frac{\alpha}{\tau_p}} 
    \approx 0.08 \Fpg
    \label{eq:rhomid-rat}
\end{equation}
where we used $\Sigma_g=\dot{M}_g/3\pi \nu$ with $\nu=\alpha h^2 r^2 \Omega$, $\Sigma_p=\Fpg\dot{M}_g/2\pi r v_r$ with $v_r\simeq 2\tau_p \eta v_K$, and a pebble-to-gas scaleheight of $H_p/H_g = \sqrt{\tau_p/\alpha}$ \citep{DubrulleEtal1995}. Therefore, the particle mass-loading is unlikely to approach unity. 

\begin{figure}
    \centering
    \includegraphics[width=88mm]{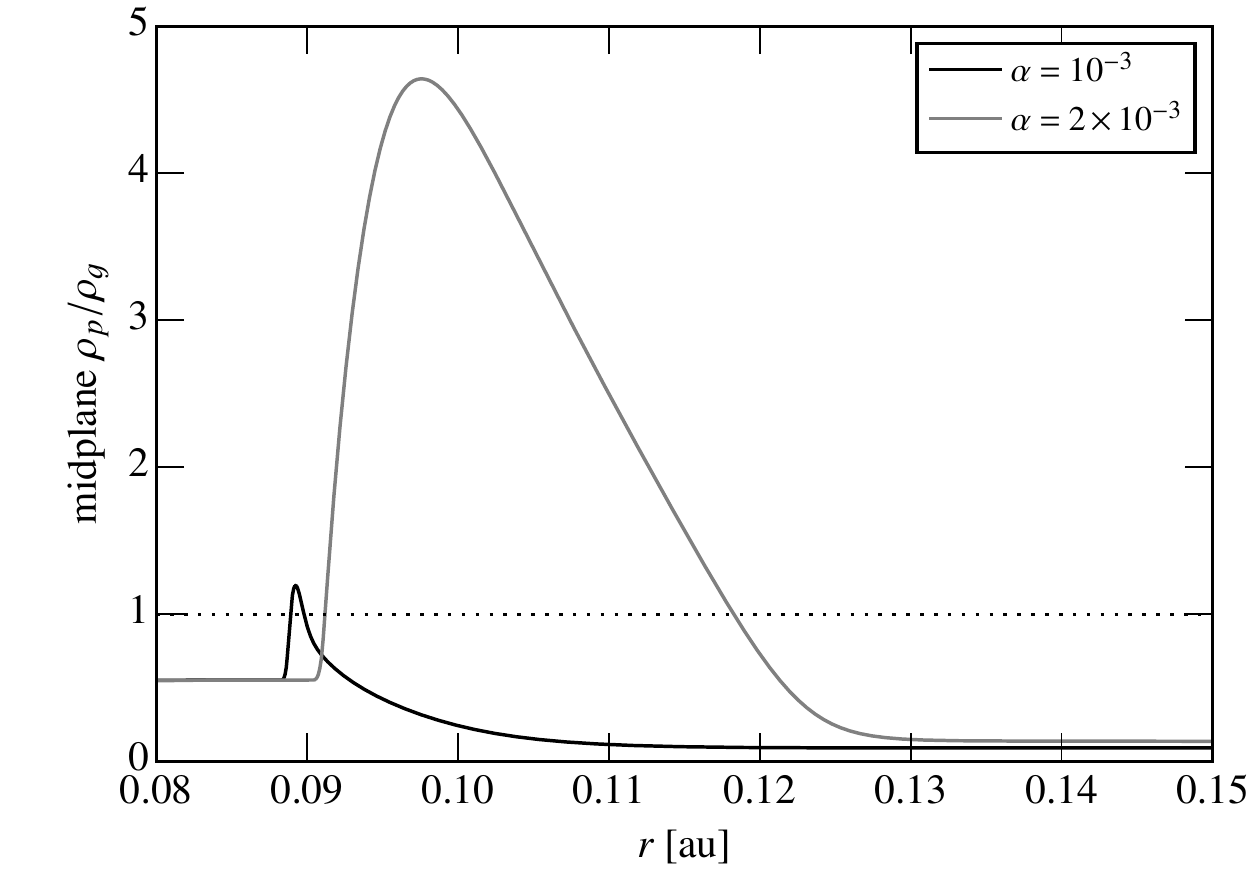}
    \caption{Midplane solids-to-gas ratio for the many seed model design of \citet{SchoonenbergOrmel2017} with $\Fpg=1.1$ for $\alpha=10^{-3}$ and $2\times10^{-3}$. The strong boost in solids-to-gas between the two runs arises from the backreaction of the solids on the gas, which reduces the radial drift of the particles resulting in a more significant pileup.
    }
    \label{fig:iceline}
\end{figure}
To further enhance $\rho_p/\rho_g$ we invoke the H$_2$O iceline, located at $\approx$$0.1\,\mathrm{au}$.  Recently, we have demonstrated that enhancements up to 10 can be attained
\citep{SchoonenbergOrmel2017}.\footnote{\citet{StevensonLunine1988} and, more recently, \citet{RosJohansen2013} already pointed to the H$_2$O iceline to trigger planetesimal formation and their subsequent growth. However, their models considered closed systems where either the particles or the gas was not removed from the system. In contrast, the steady-state model of \citet{SchoonenbergOrmel2017} features a constant inward mass flux of ice and gas.} \Fg{iceline} presents the result of the \citet{SchoonenbergOrmel2017} `many-seed' model for the iceline of \trappisti disk. 
Due to diffusion, $\hiio$ vapor is transported across the iceline where it condenses on the incoming pebbles, creating a distinct bump in the midplane pebble-to-gas ratio. For our standard parameters ($\alpha=10^{-3}$) we find that $\rho_p/\rho_g$ just exceeds unity, triggering the formation of planetesimals. The pebble-to-gas ratio increase as function of $\alpha$ for several reasons: (i) More $\hiio$ vapor is being transported at larger $\alpha$; (ii) the base $\rho_p/\rho_g$ (\eq{rhomid-rat}) increases with $\alpha$; (iii) the backreaction of the solids on the gas, which start to become important at $\rho_p/\rho_g\sim1$, reduces the particle drift, which strongly enhances the pileup effect. Finally, in \fg{iceline} increasing $\alpha$ by a factor two happens to coincide with the Epstein-Stokes drag transition, which causes another jump \citep{SchoonenbergOrmel2017}.

On the other hand, for $\alpha<10^{-3}$ pebble-to-gas ratios are likely to stay much below unity. However, disintegration of icy pebbles releases micron-size silicate grains, which can create strong pileups just interior to $r_\mathrm{ice}$ \citep{SaitoSirono2011}. \citet{IdaGuillot2016} found that the midplane mass-loading of (silicate) grains is high enough to trigger direct gravitational instability, especially at low $\alpha$. But even when particles re-coagulate their size is likely to be limited by collisional fragmentation or bouncing \citep[\eg][]{GuettlerEtal2010}, resulting in lower stopping times and dust-to-gas ratios high enough to trigger streaming instability \citep{BanzattiEtal2015,DrazkowskaEtal2016}. In those cases, planets start out rocky and the phase described in the next subsection does not exist.

\subsection{Migration interior to iceline (b)}
\label{sec:cross}
We assume that a single, dominant planetary embryo emerges from the planetesimal-pebble mix. Its growth is indeed very rapid: 
\begin{equation}
    \label{eq:tgrow}
    t_\mathrm{grow}
    = \frac{M_\mathrm{pl}}{\epsilon_\mathrm{PA} \Fpg \dot{M}_g}
    = \frac{q_\mathrm{pl} M_\star}{\epsilon_\mathrm{PA} \Fpg \dot{M}_g}
    = \frac{8\times10^3\,\mathrm{yr}}{\Fpg \epsilon_\mathrm{PA}} \left( \frac{q_\mathrm{pl}}{10^{-5}} \right)
\end{equation}
where $q_\mathrm{pl}=M_\mathrm{pl}/M_\star$ and $\epa$ is the pebble accretion efficiency. We evaluate $\epa$ in \se{eps-PA}.

Planets migrate inwards at a rate $r/t_I$ where $t_I$ is the type~I migration time:
\begin{equation}
    t_\mathrm{I} = \frac{h^2}{\gamma_I q_p q_\mathrm{gas} \Omega_K}
    = \frac{3\pi\alpha h^4}{\gamma_I q_\mathrm{pl} \dot{M}_g/M_\star}
    = 1.5\times10^5 \left( \frac{q_\mathrm{pl}}{10^{-5}} \right)^{-1} \,\mathrm{yr}
    \label{eq:typeI}
\end{equation}
where $q_\mathrm{gas}=\Sigma_g r^2/M_\star$ and $\gamma_I$ is of order unity \citep{TanakaEtal2002,KleyNelson2012}. The planet crosses the iceline when $t_\mathrm{grow} \approx \delta_\mathrm{ice} t_I$ where $\delta_\mathrm{ice} = \delta r_\mathrm{ice}/r$ is the fractional width of the iceline. We therefore find that the embryo moves interior to the iceline at a mass:
\begin{equation}
    M_\mathrm{cross} 
    = \sqrt{\frac{3\pi\alpha \delta_\mathrm{ice} \Fpg \epa}{\gamma_I}} h^2 M_\star 
    = 0.26\,M_\oplus \sqrt{\Fpg \epa}.
\end{equation}
With $\Fpg=1.1$ and $\epa=0.1$ (motivated in \se{eps-PA}) this corresponds to a Mars-mass embryo of which $\approx$$50\%$ is of icy composition. But the remainder of the accretion takes place in the interior region where pebbles are dry and the planet ends up predominantly rocky. 

Note that the value of $M_\mathrm{cross}$ depends considerably on the viscosity parameter $\alpha$. A larger $\alpha$ implies a lower gas density, suppressing migration, and a thicker ice`line' ($\delta_\mathrm{ice}$ is larger; \citealt{SchoonenbergOrmel2017}), promoting a prolonged stay in the ice-rich region.  For $\alpha\gtrsim10^{-2}$ the crossover mass may well turn out to be similar to the pebble isolation mass (see below), such that the planet will have a high H$_2$O content.

\subsection{Efficient pebble accretion (b,c)}
\label{sec:eps-PA}
Planetesimals formed by streaming instability can have sizes up to $\sim$100\,km \citep{SimonEtal2016,SchaeferEtal2017}. These planetesimals will accrete the pebbles that are drifting from the outer disk, in so-called settling interactions \citep{OrmelKlahr2010}.  This mechanism, more popularly known as pebble accretion \citep{LambrechtsJohansen2012}, is particularly attractive in case of \trappisti, because it is highly efficient: a large fraction of the pebbles are accreted.

First, consider pebble accretion exterior to the iceline. We assume that pebble accretion operates in the planar (2D) mode, which is appropriate for low-to-modest $\alpha$. Accretion of pebbles in the planar mode amounts to a rate of \citep{IdaEtal2016,Ormel2017}:
\begin{equation}
    \dot{M}_\mathrm{pl} \sim 2R_\mathrm{Hill}^2 \Omega_K \tau_p^{2/3} \Sigma_p,
    \label{eq:Mdot-pebble}
\end{equation}
where $R_\mathrm{Hill}=r(q_\mathrm{pl}/3)^{1/3}$ is the Hill radius. Since the pebble flux is $(2\pi r) v_r \Sigma_p$, pebbles are accreted at an efficiency of
\begin{equation}
    \label{eq:eps-PA}
    \epa
    \sim \frac{2}{5\cdot3^{2/3}\pi \tau_p^{1/3}} \left( \frac{q_\mathrm{pl}}{h^3} \right)^{2/3}
    = 0.1 \left( \frac{q_\mathrm{pl}}{10^{-5}} \right)^{2/3},
\end{equation}
where we inserted $v_r\approx\frac{5}{2} h^2 \tau_p \Omega_K r$ (\se{disk}).  In a more precise, N-body calculation \citep[][in prep]{LiuOrmel2017i} we find $\epsilon_\mathrm{PA}=0.25$ for $q_\mathrm{pl}=10^{-5}$. Compared to solar-type stars, pebble accretion for the \trappisti disk is particularly efficient because the disk is thin (pebbles are accreted in the 2D limit) and Hill radii are larger due to the low stellar mass. 

Next, consider pebble accretion interior to the iceline. Although the grains liberated by sublimating icy pebbles are likely to have re-coagulated, their sizes are much lower because of the silicate fragmentation threshold. Therefore, $\tau_s \ll \tau_p$ (`s' referring to silicate pebbles) and pebble accretion operates in the 3D limit \citep{IdaEtal2016,Ormel2017}:
\begin{equation}
    \label{eq:Mdot-pebble-3D}
    \dot{M}_\mathrm{pl-3D}
    \sim 6\pi R_\mathrm{Hill}^3 \tau_s \Omega \rho_s
\end{equation}
where $\rho_s$ is the midplane density of silicate pebbles. In the limit, where silicate pebbles are distributed over the entire gas scaleheight $\rho_s = \Sigma_s/2hr$ with $\Sigma_s$ the silicate surface density interior to the iceline, we obtain an efficiency of
\begin{equation}
    \label{eq:eps-PA-3D}
    \epsilon_\mathrm{PA-3D}
    = \frac{\dot{M}_\mathrm{pl-3D}}{(2\pi r)(2\eta \tau_s r\Omega) \Sigma_s}
    \sim \frac{1}{5} \frac{q_\mathrm{pl}}{h^3}
    = 0.07 \left( \frac{q_\mathrm{pl}}{10^{-5}} \right).
\end{equation}
This estimate is conservative: settling increases efficiencies by a factor $\sqrt{\tau_s/\alpha}$ and numerically-obtained rates are typically higher by a factor two.\footnote{On the other hand, when $\tau_s<\alpha$ \eq{eps-PA-3D} will be an overestimate because the particle's radial motion is then determined by the radial velocity of the gas. But since the inner disk is viscously relaxed, high $\alpha$ in turn implies larger $\tau_s$, so the situation where $\tau_s\ll\alpha$ is not easily retrieved. For example, for a silicate pebble size of 1~mm and internal density of $3~\mathrm{g~cm}^{-3}$ we obtain $\tau_s/\alpha \approx 1.1$ for our standard parameters at $r=0.05~\mathrm{au}$.} Hence, growth remains rapid; from \eq{tgrow} we obtain a growth time of $t_\mathrm{grow}\sim10^5\,\mathrm{yr}/\Fsg$, where $\Fsg$ is now the silicate-to-gas mass flux ratio. 

Our results differ in two aspects from \citet{MorbidelliEtal2015}, who also calculated pebble accretion interior and exterior to the iceline. In \citet{MorbidelliEtal2015} embryos are kept at fixed locations (no migration) and the outer embryo, growing faster, was seen to ultimately starve the inner disk from pebbles. In our scenario there is no true competition, due to the aforementioned sequential growth and migration. The second difference is that $q_\mathrm{pl}/h^3$, which enters the efficiency expressions at different powers (\eq{eps-PA} \textit{vs} \eq{eps-PA-3D}), is much larger in \trappisti. When embryos cross the iceline we have $q_\mathrm{pl}/h^3\sim0.1$, whereas \citet{MorbidelliEtal2015} starts from $q_\mathrm{pl}/h^3\sim10^{-4}$. Therefore, in \citet{MorbidelliEtal2015} the 2D-rate (exterior to the iceline) is much larger than the 3D-rates (interior), whereas in the case of \trappisti there is no significant difference.

\subsection{Pebble isolation (d)}
\label{sec:peb-iso}
Pebble accretion terminates when the gravitational feedback of the planet on the disk becomes important.  At regions where pressure maxima emerge, pebbles stop drifting. Essentially, pebble isolation describes the onset of gap opening. While the pebble isolation mass for disks around solar-type stars at 5\,au is $\approx$$20\,M_\oplus$ \citep{LambrechtsEtal2014}, it will be much smaller at the iceline of the \trappisti disk. A necessary condition for gap opening is that the Hill radius exceeds the disk scaleheight, $q_\mathrm{pl}=M_\mathrm{pl}/M_\star>h^3$ \citep{LinPapaloizou1993}; therefore:
\begin{equation}
    M_\mathrm{p,iso} 
    \sim h^3 M_\star
    = 0.72\, M_\oplus.
    \label{eq:Mp-iso}
\end{equation}
These arguments have motivated us to adopt $h\approx0.03$, but this choice is not unreasonable. The fact that Earth-mass planets naturally emerge from the pebble-driven growth scenario is a distinctive feature of the model.

After the inner-most planet first reaches the pebble isolation mass, which from the above reasoning occurs after $\sim$$2\times10^5\,\mathrm{yr}$, silicates pebbles no longer accrete on \trappisti. From that point on, the entire silicates mass reservoir -- except those grains so tiny that they follow the gas \citep{ZhuEtal2012} -- are available to make planets, resulting in a high global formation efficiency. Before isolation has been reached, the inner-most planet loses $(1-\epsilon_\mathrm{PA-3D})/\epsilon_\mathrm{PA-3D}$ pebbles for every silicate pebble it accretes. Evaluating this number at the final mass of the planet (1 Earth mass or $q_\mathrm{pl}=3.7\times10^{-5}$) we obtain that $\sim$3 Earth mass in silicates are lost. Our mechanism therefore efficiently turns solids into planets. An efficient mechanism is indeed necessary, because the initial disk contains only 11 Earth mass in rocky materials (\se{disk}).

\subsection{Migration and resonance trapping (d)--(e)}
\label{sec:res-trap}
On timescales $\sim$$t_I$ planets Type-I migrate to the disk edge ($r_c$). We assume here, for simplicity, that the migration is always inwards  -- \ie\ no special thermodynamical effects that could reverse the migration sign \citep{PaardekooperEtal2011,Benitez-LlambayEtal2015}. 
The processes illustrated in \fg{sketch}a--c then repeat until a convoy of seven planets is established.

Convergent migration of planets naturally results in resonant trapping \citep{TerquemPapaloizou2007}. For our disk model planets are likely to be trapped in the 2:1 MMR. According to \citet{OgiharaKobayashi2013} the condition to avoid trapping in the 2:1 resonance reads\footnote{Here, we evaluated Equation\,(4) of \citealt{OgiharaKobayashi2013} for $M_\mathrm{pl}=M_\oplus$ and $M_\star=0.08M_\odot$.}:
\begin{equation}
    t_I < t_\mathrm{a,crit} \approx 4\times10^3\,\mathrm{yr} \left( \frac{r}{0.1\,\mathrm{au}} \right)^{3/2}
    \label{eq:OK13}
\end{equation}
\ie we have that $t_I$ (\eq{typeI}) is too long by a factor $\approx$$10$ at 0.1\,au and the disparity only increases with lower $r$.  However, most planets are observed near the 3:2 MMR. A quantitative model to explain the settling into the 3:2 MMR resonance is beyond the scope of this \paplet but we offer several ideas for further investigation:
\begin{itemize}
    \item The \trappisti planets will form a resonant convoy, where all outer planets `push'  on the inner-most ones, effectively increasing $\gamma_I$ and decreasing the migration timescale \citep{McNeilEtal2005}. For low-$\alpha$ disk ($\alpha=10^{-4}$) gas densities are also larger by a factor 10 and we can expect planets to be moved across the 2:1.
    \item Stochastic forces, \eg triggered by density fluctuations in a magneto-rotational unstable (MRI) disk \citep{OkuzumiOrmel2013}, can move planets across resonances \citep{PaardekooperEtal2013}. Similarly, the disk accretion rate $\dot{M}_g$, assumed constant here, is likely to vary in time \citep{Hartmann2009}.  At intervals where $\dot{M}$ peaks, the condition $t_I<t_\mathrm{a,crit}$ can be met.
    \item At the time when the outer planet crosses the iceline, its period ratios with the inner planet happens to lie within the 2:1 and 3:2 resonance locations. Since $t_\mathrm{grow} \le t_I$ it is indeed plausible that the inner planet will not have migrated very far. Nevertheless, a certain level of fine-tuning is required for the planets to start within the 3:2 and 2:1 window.
\end{itemize}
Investigating each of these scenarios requires a dedicated numerical simulations. In the following, we just assume that the planets end up in MMR resonance, as shown in \fg{sketch}e.

\subsection{Disk dispersal and rebound (f)--(h)}
\begin{figure}
    \centering
    \includegraphics[width=88mm]{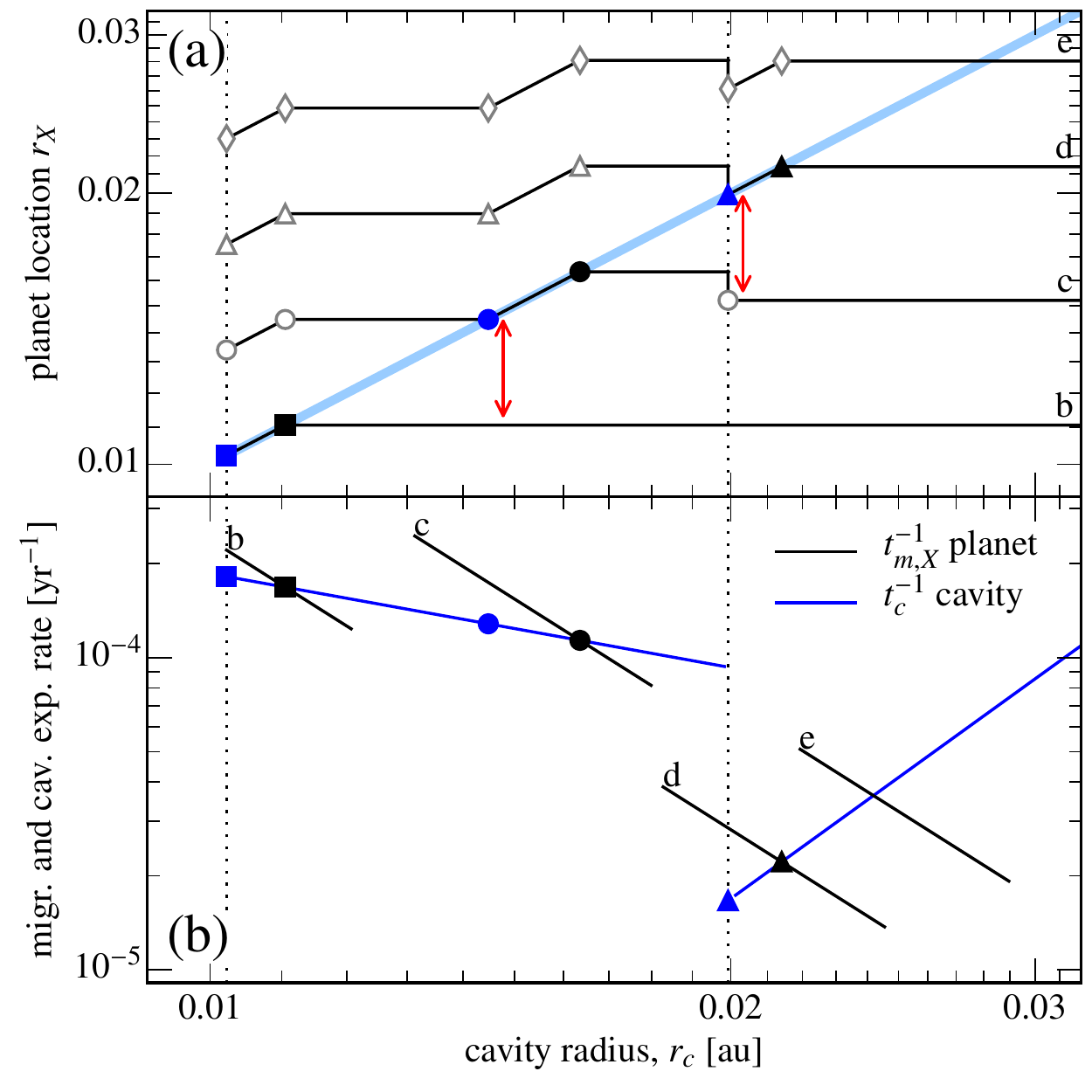}
    \caption{Resonance escape. \textit{Bottom}: one-sided migration rates $t_{m,X}^{-1}$ as function of cavity radius $r_c$ for planet $X$ (black lines; labeled) and expansion rate of the magnetospheric cavity $t_c^{-1}$ (blue lines). The form $t_c(r_c)$ is chosen to fit the orbital configuration of the \trappisti system. The jump in $t_c$ at $r_c=0.02\,\mathrm{au}$ indicates that $\dot{M}_g$ and $r_c(t)$ are temporarily constant. Blue symbols denote a planet that couples to (move synchronously with) the cavity radius. Black symbols denote the point of decoupling (the planet falls into the cavity).
    \textit{Top}: the location of the planets corresponding to $r_c(t)$. Red arrows indicate the point where the 3:2 resonance is broken, because of divergent migration.}
    \label{fig:rebound}
\end{figure}
Planet pairs b/c and c/d are presently not near the 3:2 MMR. Several mechanisms have been proposed to move planets out of resonance, \eg damping by stellar tides \citep{LithwickWu2012} or giant impacts \citep{OgiharaEtal2015}. Here we consider \textit{magnetospheric rebound} \citep{LiuEtal2017}, which relies on the outward movement of the stellar magnetospheric cavity $r_c$ during disk dispersal. In principle, the inner-most planet at $r=r_c$ is tightly coupled to the disk by strong one-sided torques \citep{LiuEtal2017}. It therefore tends to follow the expanding $r_c$. But when the cavity expansion is too rapid -- \ie when the expansion rate is higher than the planet migration rate -- the planet decouples and falls in the magnetospheric cavity. 

We have tuned the detailed behavior of $r_c(t)$ (or, equivalently, $\dot{M}(t)$ by \eq{rc}) in order to retrieve the orbital parameters of \trappisti. In \fg{rebound}a we plot the positions of planets b--e as function of the (expanding) cavity radius $r_c$ (for clarity we omit planets f, g, and h). Note that $r_c$ is used as a proxy for time. In \fg{rebound}b we plot the rate at which the cavity expands ($t_c^{-1}=\dot{r}_c/r_c$, blue curve) and the maximum one-sided planet migration rates, $t_{m,X}^{-1}$, for the planets (b, c, d, or e):
\begin{equation}
t_{\mathrm{m},X}^{-1} = 2C_\mathrm{hs} q_\mathrm{gas} \left( \frac{q_\mathrm{pl}}{h^3} \right)^{1/2} \Omega_K \propto C_X \left( \frac{r_c}{r_{c0}} \right)^{-3.5}.
    \label{eq:tmX}
\end{equation}
\citep{LiuEtal2017}.  The key point is that for any planet $t_{m,X}^{-1} \propto \dot{M}_g \propto r_c^{-3.5}$ (see \eq{rc}) and that it depends, through $C_X$, on the planet properties: more massive planets tend to have larger $C_X$ and $C_X$ also decreases with the number of planets in a resonance chain. Here we take $t_c\sim 10^4\,\mathrm{yr}$ (with a dependence on $r_c$ as in \fg{sketch}b) and $C_X=\{1.5, 4, 2, 5\}$.

Our model features a two-stage disk dispersal. First, $r_c$ doubles from $0.01$ to $0.02\,\mathrm{au}$, where it temporarily stalls (\fg{sketch}f). Planet b decouples relatively quickly from the expanding cavity front, whereas the more massive planet c couples longer to $r_c$ due to its larger $C_X$. The 3:2 resonance between planets b and c is broken because of divergent migration just after the coupling of c to the cavity radius.  

During the time when $r_c$ pauses at $0.02\,\mathrm{au}$, the accretion rate $\dot{M}_g$ is constant. We assume that this phase takes long enough for planet d to migrate inward, such that it now coincides with the cavity radius (\fg{sketch}g). Planets c, e and f--h (not shown in \fg{sketch}) follow suit as they are still part of the resonance chain.

During the second dispersal phase planet d only briefly couples to the expanding $r_c$, but enough to escape resonance with c. Afterwards, the expansion proceeds too rapid for planets to couple. For example, at the point where $r_c$ meets planet e ($r_c\approx0.028\,\mathrm{au}$) the one-sided migration rate is at least a factor three less than $\dot{r}_c$. Therefore, e and the other outer planets stay in resonance.

In the above discussion, we have referred to `resonance' as exact commensurability (nominal resonance). However, planets moving away from exact commensurability can maintain librating resonant angles; \citet{BatyginMorbidelli2013i} found that dissipative divergence in this way keeps planet in resonance in the dynamical sense. \citet{LugerEtal2017} have shown that the \trappisti planets form a complex chain of three-body resonances.\footnote{From timing analysis it is much easier to identify a three-body resonance, for which only the mean longitude are important, rather than two-body resonance, which require the arguments of periapses.} Long-term dynamical stability would be greatly promoted when the initial libration width of the resonant angles is small \citep{TamayoEtal2017} or when stellar tides play a role \citep{Papaloizou2015}.

\section{Closing remarks}
\label{sec:summary}
In this \paplet we have outlined a new framework to understand the formation of planetary systems around very low mass stars. With some modest tuning, we have succeeded in obtaining a system whose architecture reflects that of \trappisti (\fg{sketch}h). Its most radical idea is that planets assemble at a specific location -- the H$_2$O iceline. This contrasts classical models, where planets form locally, as well as population synthesis models, which do account for migration but do not (yet) provide a physical model for the initial position of planetary embryos. Our scenario is more complete, as we provide a physical model where planetary embryos (first) form.

In design, our model much resembles the inside-out formation model by \citet{ChatterjeeTan2014}, which also acknowledges the role of drifting pebbles in spawning planets at a specific location.  In \citet{ChatterjeeTan2014} and \citet{HuEtal2016} this transition occurs at the interface of an MRI-active and MRI-`dead' region, characterized by $T\sim1\,200~\mathrm{K}$, resulting in a pressure bump where pebbles stop drifting and planet formation proceeds by direct gravitational instability \citep{GoldreichWard1973}. \citet{ChatterjeeTan2014} argue that planets, once formed, experience little migration and a second planet forms in close proximity, forming a tightly-packed system consistent with the \textit{Kepler} census.

For \trappisti, however, the active:dead transition radius will lie very close to the disk inner edge, since temperatures and disk accretions rates for these tiny stars are much lower. A variant of their, as well as our, model would be to assume that pebbles drift all the way to the disk's inner edge $r_c$. Planets then grow upto their isolation mass ($\sim$$M_\oplus$) near $r_c$. However, in that case planets will be much more tightly packed, because pebbles do not stall at MMR but at the pressure bump, of which the typical distance is the gas disk scaleheight. It would be hard to produce the \trappisti architecture from such compact conditions.

Nevertheless, several parts of the proposed scenario require further investigation. We already mentioned the difficulty of avoiding the planets to get trapped in the 2:1 MMR, because of the rather low disk mass (\se{res-trap}). Another key assumption of our scenario is that one embryo-at-a-time emerges from the iceline. Simultaneous formation of multiple embryos could result in an excited system of small embryos, suppressing growth (\citealt{KretkeLevison2014}, but see \citealt{LevisonEtal2015}). Because of the short dynamical timescales, it seems viable that bodies quickly coalesce, but this complex issue -- how does streaming instability operate in the presence of planetary embryo's? -- warrants further investigation.

A prominent feature of our model is that planets mature in gas-rich disks; there is no need for a post-disk giant impact phase. Therefore, the \trappisti planets could have accreted primordial H/He atmospheres, for which there is, presently, no indication of \citep{de-WitEtal2016}.  
However, planets close to their host star receive high doses of UV and X-ray fluxes \citep{StelzerEtal2013,BourrierEtal2017}, stripping their atmospheres away efficiently \citep{LopezEtal2012}. Composition-wise, the inferred densities of $3$--$6\,\mathrm{g\,cm^{-3}}$ \citep{GillonEtal2017} rule out that these planets are entirely composed of H$_2$O. But given the error bars on the mass, planets are consistent with large ($\sim$10\%) $\hiio$ fractions -- in particular planet f -- making these planets `water worlds', even if many (Earth) oceans of $\hiio$ were lost \citep{BolmontEtal2017}. In the context of our model the maximum $\hiio$ fraction is $\approx$50\% when the planets reach their isolation mass at the $\hiio$ iceline.

The model that we outlined here can also be applied to compact systems around higher mass stars. Early- to mid-type M-stars exhibit a large concentration of planets at close distances \citep{DressingCharbonneau2015,MuirheadEtal2015}. In addition, \textit{Kepler} has found many compact systems of super-Earth planets around solar-mass stars \citep{PetiguraEtal2013}. In our model the typical planet mass scale is $\sim$$h^3 M_\star$ (\eq{Mp-iso}), which would correspond to $\sim$10\,$M_\oplus$ for solar-mass stars (still assuming $h=0.03$). The observation that fewer large (Neptune-like) planets are found around M-stars as compared to FGK-stars \citep{MuldersEtal2015} therefore agrees with one of the pillars of our scenario: that the upper mass limit of rocky planets is set by the pebble isolation mass.

The key characteristic of our model is that close-in planets are formed by tapping the pebble flux that originated from large distances ($\sim$10--100 au). 
This assumes that pebbles drifted unimpededly: the disk has to be smooth.  Also, we have argued that the natural location to rapidly form a planet is the H$_2$O iceline, which, for solar-type stars is located at $\sim$$2$--$3$\,au. 
When growth at the iceline proceeds rapidly, a $\sim$10\,$M_\oplus$ core forms a large, pre-planetary H/He-atmospheres, which can collapse into a giant planet upon reaching the critical core mass -- and a H$_2$O-enriched atmosphere would greatly accelerate this process \citep{VenturiniEtal2015}. When indeed a giant planet would form \textit{rapidly} at the snowline, it terminates the pebble flux to the inner disk, starving it from planet-building material and rendering it dry \citep{MorbidelliEtal2016}.  Hence, we expect a dichotomy: when giant planet formation fails, pebbles can drift across the iceline to aid the growth of super-Earths and mini-Neptunes. Conversely, when a giant planet forms at the iceline we expect a dearth of planetary building blocks in the inner disk.  Therefore, the close-in super-Earth population found by \textit{Kepler} and the cold Jupiter populations found chiefly by radial velocity surveys should be anti-correlated\footnote{\citet{IzidoroEtal2015} arrived at the same prediction from a different context. In their scenario, giant planets block the migration of super-Earths and mini-Neptunes, which formed even further in the outer disk. } -- a prediction that could be tested with future exoplanet surveys.

\begin{acknowledgements}
    We thank Allona Vazan, Anders Johansen, Andrius Popovas, Carsten Dominik, Jean-Michel D\'esert, Samaya Nissanke, and Sebastiaan Krijt for useful discussions and the referee for a critical report that sharpened the presentation. The authors are supported by the Netherlands Organization for Scientific Research (NWO; VIDI project 639.042.422).  
\end{acknowledgements}

\bibliographystyle{aa}
\bibliography{ads,arXiv,addbib}

\begin{thebibliography}{76}
\expandafter\ifx\csname natexlab\endcsname\relax\def\natexlab#1{#1}\fi

\bibitem[{{Banzatti} {et~al.}(2015){Banzatti}, {Pinilla}, {Ricci},
  {Pontoppidan}, {Birnstiel}, \& {Ciesla}}]{BanzattiEtal2015}
{Banzatti}, A., {Pinilla}, P., {Ricci}, L., {et~al.} 2015, \apjl, 815, L15

\bibitem[{{Batygin} \& {Morbidelli}(2013)}]{BatyginMorbidelli2013i}
{Batygin}, K. \& {Morbidelli}, A. 2013, \aj, 145, 1

\bibitem[{{Ben{\'{\i}}tez-Llambay} {et~al.}(2015){Ben{\'{\i}}tez-Llambay},
  {Masset}, {Koenigsberger}, \& {Szul{\'a}gyi}}]{Benitez-LlambayEtal2015}
{Ben{\'{\i}}tez-Llambay}, P., {Masset}, F., {Koenigsberger}, G., \&
  {Szul{\'a}gyi}, J. 2015, \nat, 520, 63

\bibitem[{{Birnstiel} {et~al.}(2012){Birnstiel}, {Andrews}, \&
  {Ercolano}}]{BirnstielEtal2012}
{Birnstiel}, T., {Andrews}, S.~M., \& {Ercolano}, B. 2012, \aap, 544, A79

\bibitem[{{Birnstiel} {et~al.}(2010){Birnstiel}, {Dullemond}, \&
  {Brauer}}]{BirnstielEtal2010i}
{Birnstiel}, T., {Dullemond}, C.~P., \& {Brauer}, F. 2010, \aap, 513, A79

\bibitem[{{Bolmont} {et~al.}(2017){Bolmont}, {Selsis}, {Owen}, {Ribas},
  {Raymond}, {Leconte}, \& {Gillon}}]{BolmontEtal2017}
{Bolmont}, E., {Selsis}, F., {Owen}, J.~E., {et~al.} 2017, \mnras, 464, 3728

\bibitem[{{Bourrier} {et~al.}(2017){Bourrier}, {Ehrenreich}, {Wheatley},
  {Bolmont}, {Gillon}, {de Wit}, {Burgasser}, {Jehin}, {Queloz}, \&
  {Triaud}}]{BourrierEtal2017}
{Bourrier}, V., {Ehrenreich}, D., {Wheatley}, P.~J., {et~al.} 2017, \aap, 599,
  L3

\bibitem[{{Carrera} {et~al.}(2015){Carrera}, {Johansen}, \&
  {Davies}}]{CarreraEtal2015}
{Carrera}, D., {Johansen}, A., \& {Davies}, M.~B. 2015, \aap, 579, A43

\bibitem[{{de Wit} {et~al.}(2016){de Wit}, {Wakeford}, {Gillon}, {Lewis},
  {Valenti}, {Demory}, {Burgasser}, {Burdanov}, {Delrez}, {Jehin}, {Lederer},
  {Queloz}, {Triaud}, \& {Van Grootel}}]{de-WitEtal2016}
{de Wit}, J., {Wakeford}, H.~R., {Gillon}, M., {et~al.} 2016, \nat, 537, 69

\bibitem[{{Dr{\c a}{\.z}kowska} {et~al.}(2016){Dr{\c a}{\.z}kowska}, {Alibert},
  \& {Moore}}]{DrazkowskaEtal2016}
{Dr{\c a}{\.z}kowska}, J., {Alibert}, Y., \& {Moore}, B. 2016, \aap, 594, A105

\bibitem[{{Dressing} \& {Charbonneau}(2015)}]{DressingCharbonneau2015}
{Dressing}, C.~D. \& {Charbonneau}, D. 2015, \apj, 807, 45

\bibitem[{{Dubrulle} {et~al.}(1995){Dubrulle}, {Morfill}, \&
  {Sterzik}}]{DubrulleEtal1995}
{Dubrulle}, B., {Morfill}, G., \& {Sterzik}, M. 1995, Icarus, 114, 237

\bibitem[{{Frank} {et~al.}(1992){Frank}, {King}, \& {Raine}}]{FrankEtal1992}
{Frank}, J., {King}, A., \& {Raine}, D. 1992, {Accretion power in
  astrophysics.}

\bibitem[{{Gillon} {et~al.}(2016){Gillon}, {Jehin}, {Lederer}, {Delrez}, {de
  Wit}, {Burdanov}, {Van Grootel}, {Burgasser}, {Triaud}, {Opitom}, {Demory},
  {Sahu}, {Bardalez Gagliuffi}, {Magain}, \& {Queloz}}]{GillonEtal2016}
{Gillon}, M., {Jehin}, E., {Lederer}, S.~M., {et~al.} 2016, \nat, 533, 221

\bibitem[{{Gillon} {et~al.}(2017){Gillon}, {Triaud}, {Demory}, {Jehin}, {Agol},
  {Deck}, {Lederer}, {de Wit}, {Burdanov}, {Ingalls}, {Bolmont}, {Leconte},
  {Raymond}, {Selsis}, {Turbet}, {Barkaoui}, {Burgasser}, {Burleigh}, {Carey},
  {Chaushev}, {Copperwheat}, {Delrez}, {Fernandes}, {Holdsworth}, {Kotze}, {Van
  Grootel}, {Almleaky}, {Benkhaldoun}, {Magain}, \& {Queloz}}]{GillonEtal2017}
{Gillon}, M., {Triaud}, A.~H.~M.~J., {Demory}, B.-O., {et~al.} 2017, \nat, 542,
  456

\bibitem[{{G{\"u}ttler} {et~al.}(2010){G{\"u}ttler}, {Blum}, {Zsom}, {Ormel},
  \& {Dullemond}}]{GuettlerEtal2010}
{G{\"u}ttler}, C., {Blum}, J., {Zsom}, A., {Ormel}, C.~W., \& {Dullemond},
  C.~P. 2010, \aap, 513, A56

\bibitem[{{Hansen} \& {Murray}(2012)}]{HansenMurray2012}
{Hansen}, B.~M.~S. \& {Murray}, N. 2012, \apj, 751, 158

\bibitem[{{Hartmann}(2009)}]{Hartmann2009}
{Hartmann}, L. 2009, {Accretion Processes in Star Formation: Second Edition}
  (Cambridge University Press)

\bibitem[{{Ida} \& {Guillot}(2016)}]{IdaGuillot2016}
{Ida}, S. \& {Guillot}, T. 2016, \aap, 596, L3

\bibitem[{{Ida} {et~al.}(2016){Ida}, {Guillot}, \& {Morbidelli}}]{IdaEtal2016}
{Ida}, S., {Guillot}, T., \& {Morbidelli}, A. 2016, \aap, 591, A72

\bibitem[{{Izidoro} {et~al.}(2015){Izidoro}, {Morbidelli}, {Raymond},
  {Hersant}, \& {Pierens}}]{IzidoroEtal2015}
{Izidoro}, A., {Morbidelli}, A., {Raymond}, S.~N., {Hersant}, F., \& {Pierens},
  A. 2015, \aap, 582, A99

\bibitem[{{Johansen} {et~al.}(2007){Johansen}, {Oishi}, {Low}, {Klahr},
  {Henning}, \& {Youdin}}]{JohansenEtal2007}
{Johansen}, A., {Oishi}, J.~S., {Low}, M., {et~al.} 2007, \nat, 448, 1022

\bibitem[{{Johansen} {et~al.}(2009){Johansen}, {Youdin}, \& {Mac
  Low}}]{JohansenEtal2009}
{Johansen}, A., {Youdin}, A., \& {Mac Low}, M. 2009, \apjl, 704, L75

\bibitem[{{Kley} \& {Nelson}(2012)}]{KleyNelson2012}
{Kley}, W. \& {Nelson}, R.~P. 2012, \araa, 50, 211

\bibitem[{{Kretke} \& {Levison}(2014)}]{KretkeLevison2014}
{Kretke}, K.~A. \& {Levison}, H.~F. 2014, \aj, 148, 109

\bibitem[{{Krijt} {et~al.}(2016){Krijt}, {Ormel}, {Dominik}, \&
  {Tielens}}]{KrijtEtal2016}
{Krijt}, S., {Ormel}, C.~W., {Dominik}, C., \& {Tielens}, A.~G.~G.~M. 2016,
  \aap, 586, A20

\bibitem[{{Lambrechts} \& {Johansen}(2012)}]{LambrechtsJohansen2012}
{Lambrechts}, M. \& {Johansen}, A. 2012, \aap, 544, A32

\bibitem[{{Lambrechts} \& {Johansen}(2014)}]{LambrechtsJohansen2014}
{Lambrechts}, M. \& {Johansen}, A. 2014, \aap, 572, A107

\bibitem[{{Lambrechts} {et~al.}(2014){Lambrechts}, {Johansen}, \&
  {Morbidelli}}]{LambrechtsEtal2014}
{Lambrechts}, M., {Johansen}, A., \& {Morbidelli}, A. 2014, \aap, 572, A35

\bibitem[{{Lee} \& {Peale}(2002)}]{LeePeale2002}
{Lee}, M.~H. \& {Peale}, S.~J. 2002, \apj, 567, 596

\bibitem[{{Levison} {et~al.}(2015){Levison}, {Kretke}, \&
  {Duncan}}]{LevisonEtal2015}
{Levison}, H.~F., {Kretke}, K.~A., \& {Duncan}, M.~J. 2015, \nat, 524, 322

\bibitem[{{Lin} \& {Papaloizou}(1993)}]{LinPapaloizou1993}
{Lin}, D.~N.~C. \& {Papaloizou}, J.~C.~B. 1993, in Protostars and Planets III,
  ed. E.~H. {Levy} \& J.~I. {Lunine}, 749--835

\bibitem[{{Lithwick} \& {Wu}(2012)}]{LithwickWu2012}
{Lithwick}, Y. \& {Wu}, Y. 2012, \apjl, 756, L11

\bibitem[{{Liu} \& {Ormel}(2017)}]{LiuOrmel2017i}
{Liu}, B. \& {Ormel}, C.~W. 2017, in prep

\bibitem[{{Liu} {et~al.}(2017){Liu}, {Ormel}, \& {Lin}}]{LiuEtal2017}
{Liu}, B., {Ormel}, C.~W., \& {Lin}, D.~N.~C. 2017, ArXiv e-prints:1702.02059

\bibitem[{{Lodders}(2003)}]{Lodders2003}
{Lodders}, K. 2003, \apj, 591, 1220

\bibitem[{{Lopez} {et~al.}(2012){Lopez}, {Fortney}, \&
  {Miller}}]{LopezEtal2012}
{Lopez}, E.~D., {Fortney}, J.~J., \& {Miller}, N. 2012, \apj, 761, 59

\bibitem[{{Luger} {et~al.}(2017){Luger}, {Sestovic}, {Kruse}, {Grimm},
  {Demory}, {Agol}, {Bolmont}, {Fabrycky}, {Fernandes}, {Van Grootel},
  {Burgasser}, {Gillon}, {Ingalls}, {Jehin}, {Raymond}, {Selsis}, {Triaud},
  {Barclay}, {Barentsen}, {Delrez}, {de Wit}, {Foreman-Mackey}, {Holdsworth},
  {Leconte}, {Lederer}, {Turbet}, {Almleaky}, {Benkhaldoun}, {Magain},
  {Morris}, {Heng}, \& {Queloz}}]{LugerEtal2017}
{Luger}, R., {Sestovic}, M., {Kruse}, E., {et~al.} 2017, ArXiv
  e-prints:1703.04166

\bibitem[{{Lynden-Bell} \& {Pringle}(1974)}]{Lynden-BellPringle1974}
{Lynden-Bell}, D. \& {Pringle}, J.~E. 1974, \mnras, 168, 603

\bibitem[{{Manara} {et~al.}(2015){Manara}, {Testi}, {Natta}, \&
  {Alcal{\'a}}}]{ManaraEtal2015}
{Manara}, C.~F., {Testi}, L., {Natta}, A., \& {Alcal{\'a}}, J.~M. 2015, \aap,
  579, A66

\bibitem[{{McNeil} {et~al.}(2005){McNeil}, {Duncan}, \&
  {Levison}}]{McNeilEtal2005}
{McNeil}, D., {Duncan}, M., \& {Levison}, H.~F. 2005, \aj, 130, 2884

\bibitem[{{Morbidelli} {et~al.}(2016){Morbidelli}, {Bitsch}, {Crida},
  {Gounelle}, {Guillot}, {Jacobson}, {Johansen}, {Lambrechts}, \&
  {Lega}}]{MorbidelliEtal2016}
{Morbidelli}, A., {Bitsch}, B., {Crida}, A., {et~al.} 2016, \icarus, 267, 368

\bibitem[{{Morbidelli} {et~al.}(2015){Morbidelli}, {Lambrechts}, {Jacobson}, \&
  {Bitsch}}]{MorbidelliEtal2015}
{Morbidelli}, A., {Lambrechts}, M., {Jacobson}, S., \& {Bitsch}, B. 2015,
  \icarus, 258, 418

\bibitem[{{Mulders} \& {Dominik}(2012)}]{MuldersDominik2012}
{Mulders}, G.~D. \& {Dominik}, C. 2012, \aap, 539, A9

\bibitem[{{Mulders} {et~al.}(2015){Mulders}, {Pascucci}, \&
  {Apai}}]{MuldersEtal2015}
{Mulders}, G.~D., {Pascucci}, I., \& {Apai}, D. 2015, \apj, 814, 130

\bibitem[{{Nakagawa} {et~al.}(1986){Nakagawa}, {Sekiya}, \&
  {Hayashi}}]{NakagawaEtal1986}
{Nakagawa}, Y., {Sekiya}, M., \& {Hayashi}, C. 1986, Icarus, 67, 375

\bibitem[{{Ogihara} \& {Kobayashi}(2013)}]{OgiharaKobayashi2013}
{Ogihara}, M. \& {Kobayashi}, H. 2013, \apj, 775, 34

\bibitem[{{Ogihara} {et~al.}(2015){Ogihara}, {Morbidelli}, \&
  {Guillot}}]{OgiharaEtal2015}
{Ogihara}, M., {Morbidelli}, A., \& {Guillot}, T. 2015, \aap, 578, A36

\bibitem[{{Okuzumi} \& {Ormel}(2013)}]{OkuzumiOrmel2013}
{Okuzumi}, S. \& {Ormel}, C.~W. 2013, \apj, 771, 43

\bibitem[{{Ormel}(2017)}]{Ormel2017}
{Ormel}, C.~W. 2017, in Formation, Evolution, and Dynamics of Young Solar
  Systems, ed. {{Gressel}, O. \& {Pessah}, M.}, Proceedings of the Sant Cugat
  Forum on Astrophysics (Springer)

\bibitem[{{Ormel} \& {Klahr}(2010)}]{OrmelKlahr2010}
{Ormel}, C.~W. \& {Klahr}, H.~H. 2010, \aap, 520, A43

\bibitem[{{Paardekooper} {et~al.}(2011){Paardekooper}, {Baruteau}, \&
  {Kley}}]{PaardekooperEtal2011}
{Paardekooper}, S.-J., {Baruteau}, C., \& {Kley}, W. 2011, \mnras, 410, 293

\bibitem[{{Paardekooper} {et~al.}(2013){Paardekooper}, {Rein}, \&
  {Kley}}]{PaardekooperEtal2013}
{Paardekooper}, S.-J., {Rein}, H., \& {Kley}, W. 2013, \mnras, 434, 3018

\bibitem[{{Papaloizou}(2015)}]{Papaloizou2015}
{Papaloizou}, J.~C.~B. 2015, International Journal of Astrobiology, 14, 291

\bibitem[{{P{\'e}rez} {et~al.}(2015){P{\'e}rez}, {Chandler}, {Isella},
  {Carpenter}, {Andrews}, {Calvet}, {Corder}, {Deller}, {Dullemond}, {Greaves},
  {Harris}, {Henning}, {Kwon}, {Lazio}, {Linz}, {Mundy}, {Ricci}, {Sargent},
  {Storm}, {Tazzari}, {Testi}, \& {Wilner}}]{PerezEtal2015}
{P{\'e}rez}, L.~M., {Chandler}, C.~J., {Isella}, A., {et~al.} 2015, \apj, 813,
  41

\bibitem[{{Petigura} {et~al.}(2013){Petigura}, {Marcy}, \&
  {Howard}}]{PetiguraEtal2013}
{Petigura}, E.~A., {Marcy}, G.~W., \& {Howard}, A.~W. 2013, \apj, 770, 69

\bibitem[{{Rafikov} \& {De Colle}(2006)}]{RafikovDe-Colle2006}
{Rafikov}, R.~R. \& {De Colle}, F. 2006, \apj, 646, 275

\bibitem[{{Reiners} {et~al.}(2009){Reiners}, {Basri}, \&
  {Christensen}}]{ReinersEtal2009}
{Reiners}, A., {Basri}, G., \& {Christensen}, U.~R. 2009, \apj, 697, 373

\bibitem[{{Ricci} {et~al.}(2012){Ricci}, {Testi}, {Natta}, {Scholz}, \& {de
  Gregorio-Monsalvo}}]{RicciEtal2012}
{Ricci}, L., {Testi}, L., {Natta}, A., {Scholz}, A., \& {de Gregorio-Monsalvo},
  I. 2012, \apjl, 761, L20

\bibitem[{{Ros} \& {Johansen}(2013)}]{RosJohansen2013}
{Ros}, K. \& {Johansen}, A. 2013, \aap, 552, A137

\bibitem[{{Saito} \& {Sirono}(2011)}]{SaitoSirono2011}
{Saito}, E. \& {Sirono}, S.-i. 2011, \apj, 728, 20

\bibitem[{{Sato} {et~al.}(2016){Sato}, {Okuzumi}, \& {Ida}}]{SatoEtal2016}
{Sato}, T., {Okuzumi}, S., \& {Ida}, S. 2016, \aap, 589, A15

\bibitem[{{Sch{\"a}fer} {et~al.}(2017){Sch{\"a}fer}, {Yang}, \&
  {Johansen}}]{SchaeferEtal2017}
{Sch{\"a}fer}, U., {Yang}, C.-C., \& {Johansen}, A. 2017, \aap, 597, A69

\bibitem[{{Schoonenberg} \& {Ormel}(2017)}]{SchoonenbergOrmel2017}
{Schoonenberg}, D. \& {Ormel}, C.~W. 2017, ArXiv e-prints:1702.02151

\bibitem[{{Simon} {et~al.}(2016){Simon}, {Armitage}, {Li}, \&
  {Youdin}}]{SimonEtal2016}
{Simon}, J.~B., {Armitage}, P.~J., {Li}, R., \& {Youdin}, A.~N. 2016, \apj,
  822, 55

\bibitem[{{Stelzer} {et~al.}(2013){Stelzer}, {Marino}, {Micela},
  {L{\'o}pez-Santiago}, \& {Liefke}}]{StelzerEtal2013}
{Stelzer}, B., {Marino}, A., {Micela}, G., {L{\'o}pez-Santiago}, J., \&
  {Liefke}, C. 2013, \mnras, 431, 2063

\bibitem[{{Stevenson} \& {Lunine}(1988)}]{StevensonLunine1988}
{Stevenson}, D.~J. \& {Lunine}, J.~I. 1988, \icarus, 75, 146

\bibitem[{{Tamayo} {et~al.}(2017){Tamayo}, {Rein}, {Petrovich}, \&
  {Murray}}]{TamayoEtal2017}
{Tamayo}, D., {Rein}, H., {Petrovich}, C., \& {Murray}, N. 2017, ArXiv
  e-prints:1704.02957

\bibitem[{{Tanaka} {et~al.}(2002){Tanaka}, {Takeuchi}, \&
  {Ward}}]{TanakaEtal2002}
{Tanaka}, H., {Takeuchi}, T., \& {Ward}, W.~R. 2002, \apj, 565, 1257

\bibitem[{{Terquem} \& {Papaloizou}(2007)}]{TerquemPapaloizou2007}
{Terquem}, C. \& {Papaloizou}, J.~C.~B. 2007, \apj, 654, 1110

\bibitem[{{Testi} {et~al.}(2014){Testi}, {Birnstiel}, {Ricci}, {Andrews},
  {Blum}, {Carpenter}, {Dominik}, {Isella}, {Natta}, {Williams}, \&
  {Wilner}}]{TestiEtal2014}
{Testi}, L., {Birnstiel}, T., {Ricci}, L., {et~al.} 2014, Protostars and
  Planets VI, 339

\bibitem[{{Venturini} {et~al.}(2015){Venturini}, {Alibert}, {Benz}, \&
  {Ikoma}}]{VenturiniEtal2015}
{Venturini}, J., {Alibert}, Y., {Benz}, W., \& {Ikoma}, M. 2015, \aap, 576,
  A114

\bibitem[{{Weidenschilling}(1977)}]{Weidenschilling1977}
{Weidenschilling}, S.~J. 1977, \mnras, 180, 57

\bibitem[{{Yang} {et~al.}(2016){Yang}, {Johansen}, \& {Carrera}}]{YangEtal2016}
{Yang}, C.-C., {Johansen}, A., \& {Carrera}, D. 2016, ArXiv e-prints:1611.07014

\bibitem[{{Youdin} \& {Goodman}(2005)}]{YoudinGoodman2005}
{Youdin}, A.~N. \& {Goodman}, J. 2005, \apj, 620, 459

\bibitem[{{Zhu} {et~al.}(2012){Zhu}, {Nelson}, {Dong}, {Espaillat}, \&
  {Hartmann}}]{ZhuEtal2012}
{Zhu}, Z., {Nelson}, R.~P., {Dong}, R., {Espaillat}, C., \& {Hartmann}, L.
  2012, \apj, 755, 6

\end{thebibliography}

\end{document}